\documentclass[conference]{IEEEtran}
\ifCLASSINFOpdf
\else
\fi
\usepackage[cmex10]{amsmath}
\usepackage{amssymb}
\usepackage{xfrac}

\usepackage{tikz}
\usetikzlibrary{positioning,arrows,shapes.geometric}
\usetikzlibrary{patterns}
\tikzstyle{int} = [draw, fill=blue!20, minimum size=2em]
\tikzstyle{init} = [pin edge={to-,thin,black}] 
\tikzstyle{block} = [draw, fill=blue!20, rectangle, 
    minimum height=3em, minimum width=4em]
\tikzstyle{sum} = [draw, fill=blue!20, circle, node distance=1cm]
\tikzstyle{input} = [coordinate]
\tikzstyle{output} = [coordinate]
\tikzstyle{pinstyle} = [pin edge={to-,thin,black}]

\usepackage{bigstrut}
\usepackage{pgfplots}

\newcommand{\originalsystem}[1][]{ 
\scalebox{0.8} {
	\begin{tikzpicture}[auto, node distance=2.5cm, >= latex']
    		\node [input, name=input] {};
    		\node [block, right of=input] (system) {$f$};
    		\node [output, right of=system] (output) {};

    		\draw [draw,->] (input) -- node {$ \upsilon_2^i $} (system);
    		\draw [->] (system) -- node [name=y] {$ \dfrac{\partial{\tilde{\lambda}_2}}{\partial{p_i}} $}(output);
	
	\end{tikzpicture}
	}
}

\newcommand{\modifiedsystem}[1][]{
\scalebox{0.8} {
	\begin{tikzpicture}[auto, node distance=2.5 cm, scale = 1, every node/.style={transform shape}]     
		\node [input, name=input] {};
    		\node [block, right of=input] (fail) {$failure$};
    		\node [block, right of=fail] (system) {$f$};
    		\node [block, right of=system] (noise) {$noise$};
    		\node [output, right of=noise] (output) {};

    		\draw [draw,->] (input) -- node {$ \upsilon_2^i $} (fail);
   		\draw [->] (fail) -- node [name=y1] {$  $}(system);
    		\draw [->] (system) --node [name=y2] {$  $}(noise);
    		\draw [->] (noise) --node [name=y4] {$ \dfrac{\partial{\tilde{\lambda}_2}}{\partial{p_i}} $} (output);
	\end{tikzpicture}
	}
}

\begin{document}
%
\title{Analysis of the Effects of Failure and Noise in the Distributed Connectivity Maintenance of a Multi-robot System}

\author{\IEEEauthorblockN{Vin\'{i}cius A. Battagello}
\IEEEauthorblockA{Postgraduate Program\\ Computing and Electronical Engineering\\
Technological Institute of Aeronautics (ITA)\\
S\~{a}o Jos\'{e} dos Campos, SP -- Brazil\\
Email: batta@ita.br}
\and
\IEEEauthorblockN{Carlos H. C. Ribeiro}
\IEEEauthorblockA{Technological Institute of Aeronautics (ITA)\\
S\~{a}o Jos\'{e} dos Campos, SP -- Brazil\\
Email: carlos@ita.br}
}


%


\maketitle

\begin{abstract}
To perform cooperative tasks in a decentralized manner, multi-robot systems are often required to communicate with each other. Therefore, maintaining the communication graph connectivity is a fundamental issue when roaming a territory with obstacles. However, when dealing with real-robot systems, several sources of data corruption can appear in the agent interaction. In this paper, the effects of failure and noise in the communication between agents are analyzed upon a connectivity maintenance control strategy. The results show that the connectivity strategy is resilient to the negative effects of such disturbances under realistic settings that consider a bandwidth limit for the control effort. This opens the perspective of applying the connectivity maintenance strategy in adaptive schemes that consider, for instance, autonomous adaptation to constraints other than connectivity itself, e.g. communication efficiency and energy harvesting.
\end{abstract}


%
\IEEEpeerreviewmaketitle

\section{Introduction}
In this paper, the effects of different disturbance types (communication failures and noise) upon the control strategy introduced in \cite{sabattini2011} are analyzed and its impact upon the connectivity maintenance of multi-robot systems is evaluated. As this is a decentralized strategy that requires local communication between agents, the main requirement is to keep the agents always connected during communication.

The strategy considered here is a representative of the so-called \textit{global} connectivity techniques, which aim at maintaining a path between any pair of nodes (\emph{i.e.} robots), with a possible elimination of redundant links and creation of new links as needed. This is in contrast with \textit{local} connectivity techniques such as \cite{JiEgersted2007}, \cite{hsieh2008} and \cite{cao2010}, that guarantee that once a communication link between a pair of agents is active at time $t = 0$, it will be active $\forall t > 0$, thus establishing an initial connection path from the outset. From the Engineering point of view, however, imposing the maintenance of each single communication link is very costly, and a more intelligent scheme that removes redundant links and generate additional links as required can be more convenient. Indeed, information exchange among all the agents is guaranteed under a \textit{global} connectivity of the communication graph. For example, in \cite{hollinger2010} the idea of periodic connectivity, in which the network regains connectivity at predefined intervals, is introduced.

The control strategy introduced in \cite{sabattini2011} ensures that \textit{global} connectivity is maintained in a system of agents. This approach is inspired in the strategy previously defined in \cite{yang10} and introduces a way of obtaining, in a distributed way, the value of the second smallest eigenvalue of the Laplacian matrix, that, as shown in \cite{fiedler73}, measures the connectivity of a graph. Regarding other estimate procedures that can be found in \cite{degenaro06} and \cite{zavlanos09}, one of the principal advantages of the method described in \cite{sabattini2011} is that it provides, besides the Fiedler eigenvalue, estimations of its own gradients that are useful in real-time computations, as will be verified below.

 

\section{Background on Graph Theory}
The connectivity scheme considered here is based on graph-theoretical considerations that are outlined in this section. Further background details can be found in \cite{godsil2001}. 

Given $N$ mobile robots, the instantaneous communication links among them are modeled as edges in an undirected graph where each robot corresponds to a node. Communication is assumed to be local, in the sense that each robot communicates only with a topological neighbourhood formed by nearby robots. Thus, $\mathcal{N}_i$ is defined as the neighborhood of the $i$-th robot, i.e. the set of robots that can exchange information with it. The complete communication graph is represented by the adjacency matrix $A \in \mathbb{R}^{N \times N}$, where each $a_{ij}$ is defined as the weight of the edge between robots $i$ and $j$, and is positive if $j \in \mathcal{N}_i$, zero otherwise. This value needs to be computed in order to allow the local Laplacian evaluation, as we'll see next. As undirected graphs are being considered, $a_{ij} = a_{ji}$. 

Finally, consider the Laplacian matrix of the graph, defined as $L = D - A$ where $\mathit{D} = diag(\lbrace d_i \rbrace)$, and $d_i = \sum\limits_{j=1}^{N}a_{ij}$ is the degree of the $i$-th node of the graph. $L$ holds some important properties, among which a remarkable relationship between its eigenvalues and the graph connectivity. Namely, let $\lambda_i$, $i = 1, \ldots, N$ be the eigenvalues of $L$. Then
\begin{itemize}
	\item The eigenvalues can be ordered such that
				\begin{equation}	 \label{eq:eigenvalues_relation}
					0 = \lambda_1 \leq \lambda_2 \leq  \ldots \leq \lambda_N
				\end{equation}
	\item $\lambda_2 > 0$ if and only if the graph is connected: $\lambda_2$ is then defined as the algebraic connectivity of the graph.
\end{itemize}

This means that any procedure that keeps the second eigenvalue $\lambda_2$ of the communication graph at positive values guarantees graph connectivity, \emph{i.e.}, guarantees a communication path between any pair of nodes.

\section{Connectivity Maintenance}
In this section a control strategy assuming that each agent can compute the actual value of $\lambda_2$ is first summarized. This hypothesis will be removed in the sequence, with the description of the distributed procedure.

\subsection{Centralized Connectivity Maintenance Control Strategy} \label{centralized_strategy}
For the sake of clarity, a brief synopsis of the estimation procedure introduced in \cite{yang10} will now be made. The estimation of $\lambda_2$ is computed by each agent through an estimate of the corresponding eigenvector $\upsilon_2$. 

Considering a group of $N$ single integrator agents, let $p_i \in \mathbb{R}^m$ be the state vector describing the position of the $i$-th agent and $u_i$ be the its control input. Then:
\begin{equation} \label{eq:control_law0}
	\dot{p}_i = u_i^c
\end{equation}
where $p = \left[ p_1^T \ldots p_N^T \right]$ and $u_i^c$ is defined as:
\begin{equation}
	u_i^c = csch^2(\lambda_2 - \epsilon).\dfrac{\partial{\lambda}_2}{\partial{p}_i}
\end{equation}
This control law ensures connectivity in a centralized framework, supposing each agent can compute the real value of the algebraic connectivity of the system. Subsequently, this assertion will be removed with the decentralized estimation procedure in Section \ref{decentralized_strategy}.

The maximum communication range for each agent is denoted by $R$, and the $j$-th agent is inside $\mathcal{N}_i$ if $\vert\vert p_i - p_j\vert\vert \leq R$. 
In order to implement the estimation procedure in a decentralized manner, a brief overview  of the estimation procedure proposed by \cite{sabattini2011} is presented in the next subsection. Specifically, the estimation of $\lambda_2$ is computed through the estimation  of the corresponding eigenvector $\upsilon_2$.

According to \cite{yang10}, $\lambda_2$ can be computed in a centralized framework as:
\begin{equation} \label{eq:dlambda2_descentralizado}
	\dfrac{\partial{\lambda_2}}{\partial{p_i}} = {\sum\limits_{j \epsilon N_i}-a_{ij}\left( \upsilon_2^i - \upsilon_2^j  \right)^2.\frac{p_i - p_j}{\sigma^2}}
\end{equation}
where the edge-weights $a_{ij}$ are defined as in Eq. \eqref{eq:aij}:
\begin{equation} \label{eq:aij}
	a_{ij} = \begin{cases}
	\ e^{\sfrac{-(||p_i - p_j||)^2}{(2.\sigma^2)}} & if \ \vert\vert p_i-p_j \vert\vert \leq R \\
	 0 \ , 	& otherwise
	\end{cases} 
\end{equation}
where the scalar parameter $\sigma$ is chosen to satisfy the boundary condition $e^{\sfrac{-(R^2)}{(2.\sigma^2)}} = \Delta$, in which $\Delta$ is a small defined threshold.

\subsection{Decentralized Connectivity Maintenance Control Strategy} \label{decentralized_strategy}
The control law implemented in this work was proposed in \cite{sabattini2011} and extends the one introduced in Eq. \eqref{eq:control_law0} by adding a bounded control term $u^d$ to obtain some desired behavior:
\begin{equation} \label{eq:control_law_decentralized}
	\dot{p}_i = u_i^c + u_i^d
\end{equation}
where $u_i^c$ is given by Eq. \eqref{eq:u^c_final}: 
\begin{equation} \label{eq:u^c_final}
	u_i^c = csch^2(\lambda_2 - \epsilon).\dfrac{\partial{\tilde{\lambda}}_2}{\partial{p}_i}
\end{equation}

To implement the strategy described in \cite{sabattini2012} in a decentralized way, each agent must compute an estimate of a component of the eigenvector $\upsilon_2$, given by $\tilde{\upsilon}_2^i$. Let $\tilde{\upsilon}_2 = \left[\tilde{\upsilon}_2^1 \ldots \tilde{\upsilon}_2^N \right]^T$ and $\tilde{\lambda}_2$ be the second smallest eigenvalue that the Laplacian matrix would take if $\tilde{\upsilon}_2$ were the corresponding eigenvector.
So $\tilde{\lambda}_2$ can be computed in the following way:
\begin{equation}	\label{eq:lambda2_decentralized}
	\tilde{\lambda}_2 = \dfrac{k_3}{k_2}.\left[1 - Ave({(\tilde{\upsilon}_2^i)^2   }) \right]
\end{equation}
According to the procedure given in \cite{sabattini2012}, each agent can compute its own estimate of $\lambda_2$ using:
\begin{equation} \label{eq:dlambda2_decentralized}
	\dfrac{\partial{\tilde{\lambda}_2}}{\partial{p_i}} = f\left( \tilde{\upsilon}_2^i, \tilde{\upsilon}_2^j \right) = \sum\limits_{j \epsilon N_i}-a_{ij}\left( \tilde{\upsilon}_2^i - \tilde{\upsilon}_2^j  \right)^2.\frac{p_i - p_j}{\sigma^2}
\end{equation}

However, the actual value of $\tilde{\lambda}_2$ cannot be computed by each agent, because actually, the real value of $Ave(\lbrace(\tilde{\upsilon}_2^i)\rbrace)$ cannot be calculated in a distributed way. Nevertheless, an estimate of this average is available to each agent (see \cite{yang10} for further details).

As shown in \cite{sabattini2012}, $\lambda_2^i$ is a good estimate of both $\lambda_2$ and $\tilde{\lambda}_2$. In other words, it has been shown that, given a positive $\Xi$  value, it is always possible to guarantee that the absolute difference between the real (and distributedly unreachable) value assumed by $\lambda_2$ and its estimate obtained by agent $i$ (given by $\lambda_2^i$) is bounded from $0$ by $\Xi$ for every agent. Additionally, it has also been shown that, given another positive $\Xi^{\prime}$ value, it is always possible to ensure that the absolute difference between the second smallest eigenvalue of $L$ (that can be computed distributedly and denoted by $\tilde{\lambda}_2$) and its estimate obtained by agent $i$ (given by $\lambda_2^i$) is bounded from $0$ by $\Xi^{\prime}$.

That is, it can be concluded then that, given a positive integer $\Xi^{\prime\prime}$ (given by $\Xi + \Xi^\prime$), the absolute difference between the real (and distributedly unreachable) value assumed by $\lambda_2$ and the second smallest eigenvalue of $L$ (which can be computed distributedly) is bounded from $0$ by $\Xi^{\prime\prime}$. Put differently, each agent is able to (locally) compute using $\tilde{\lambda}_2$ instead of $\lambda_2$ and still be able to obtain a valid measure of the (global) system connectivity.

Even though the actual value of $\tilde{\lambda}_2$ is not available to each agent, it is worth noting that the partial derivatives of $\tilde{\lambda}_2$ can be implemented in a decentralized manner. This way, each robot can locally estimate the value of $\lambda_2$ using Eq. \eqref{eq:lambda2_decentralized}.

\subsubsection{Rendezvous and Formation Control Strategy}
This section will summarize the main notions of the control strategy studied in this work to address the connectivity maintenance within rendezvous and formation control problems.

The edge-weights $\bar{a}_{ij}$ were redefined so that the connectivity control action was implemented in its vector form:
\begin{equation} \label{eq:aij_final}
	\bar{a}_{ij} (\lambda_2^i) = \gamma_i.\textit{csch}^2(\lambda_2^i - \bar{\epsilon}).\dfrac{1}{\sigma^2}.\left( \tilde{\upsilon}_2^i-\tilde{\upsilon}_2^j \right)^2.a_{ij}
\end{equation}
Additionally, a modified degree matrix of the graph was introduced as $\bar{D} = diag(\lbrace \bar{d}_i\rbrace)$, where $\bar{d}_i = \sum\limits_{j=1}^{N}\bar{a}_{ij}$, i.e. represents the degree of the $i$-th node of the graph. The modified laplacian matrix of the graph, then, is redefined accordingly as $\bar{L} = \bar{D} - \bar{A}$, and the control law of the system assumes the form:
\begin{equation} \label{eq:leidecontrole_descentralizado7_final}
	\dot{p} = -\bar{L}.p + u^d 
\end{equation}
where $u^d$ is a vector containing the control laws for each one of the $N$ agents. 
\subsubsection*{Consensus-Based Rendezvous}
According to \cite{Saber03consensusproblems}, the following consensus-based control law guarantees the convergence of the robots to the same position if the communication graph is connected:
\begin{equation}	 \label{eq:ud_rendezvous}
	u^d = -L*p
\end{equation}
\subsubsection*{Consensus-Based Formation Control} 
This is the final form of the implemented algorithm. Instead of migrating to the same position, the agents assume a specific formation. As stated in \cite{fax04}, the following control strategy can be used, by adding a bias term $b_i(p)$ to the control law in Eq. \eqref{eq:ud_rendezvous}:
\begin{equation} \label{eq:ud_formationcontrol}
	u^d = -L*p + b_i(p)
\end{equation}
where
\begin{equation} \label{eq:b_formationcontrol}
	b_i(p) = \begin{cases}
	\ \sum\limits_{j \epsilon N_i}\left( 1+\bar{a}_{ij}(\lambda_2^i) \right).\left( \bar{p}_i - \bar{p}_j \right) &, \ if \ \lambda_2^i > k.\tilde{\epsilon}  \\
	 \ \sum\limits_{j \epsilon N_i}\left( 1+\bar{a}_{ij}(k.\tilde{\epsilon}) \right).\left( \bar{p}_i - \bar{p}_j \right) &, \ otherwise
	\end{cases} 
\end{equation}
for some $k > 1$, where $\bar{p}_i$ represents the desired relative position for robot $i$ in the formation. This way, when the estimate of the algebraic connectivity is sufficiently greater than $\tilde{\epsilon}$ (i.e. $\lambda_2^i > k.\tilde{\epsilon}$), the bias term is computed with Laplacian matrix $\tilde{L} = \bar{L} + L_{*}$. When the value of the estimate of the algebraic connectivity falls and approaches $\tilde{\epsilon}$, the bias term of the second case in Eq. \eqref{eq:b_formationcontrol} ensures $u^d$ is bounded and guarantees the connectivity maintenance among the agents.  

\section{The Disturbance Model}

\begin{figure}
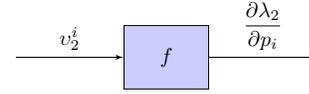

\centering
\originalsystem{}
\caption{$Block \ Diagram \ of \ the \ Original \ Model$} \label{fig:blockdiagram_originalsystem}
\end{figure}

Each agent $i$ computes its own $\upsilon_2$ estimate, given by $\upsilon_2^i$, and from it obtains the spatial variation of the corresponding eigenvalue, given by $\sfrac{{\partial{\tilde{\lambda}_2}}}{\partial{p_i}}$ as explained in Section \ref{decentralized_strategy}. This estimation system can be represented by the block diagram in Fig. \ref{fig:blockdiagram_originalsystem}, where $f(\cdot)$ is given by Eq. \eqref{eq:dlambda2_decentralized}.

In order to maintain the communication path between every pair of robots, each robot should be able to estimate effectively the global connectivity of the system only with local information. However, both communication failures and data corruption (noise) coexist in a realistic multi-robot communication system that is expected to keep itself connected, even in the presence of such disturbances.

\begin{figure}
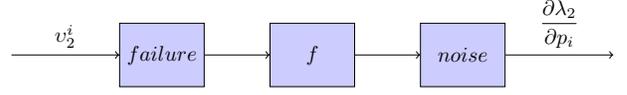

\centering
\modifiedsystem{}
\caption{\textit{Block Diagram of the Model with Disturbances}} \label{fig:blockdiagram_modifiedsystem}
\end{figure}

The model proposed in this work adds \textit{failure} and \textit{noise} as sources of corruption into the arrangement proposed in \cite{sabattini2012} to study its effects on the final system performance. This situation can then be represented by the block diagram in Fig. \ref{fig:blockdiagram_modifiedsystem}.

\subsection{Modelling failures}
The local estimation of $\upsilon_2$ made by each agent is a critical process to keep the system connected. Errors in this estimate automatically imply incorrect values of $\tilde{\lambda}_2$, which may result in $\lambda_2 = 0$. Given $N$ agents interacting in an unknown environment, failures in data reception are considered. Thus, let $\upsilon_2^{i^\prime}$ be the $\upsilon_2$ estimate made by agent $i$ in the presence of a communication failure. In this case, it is assumed that the agent stipulates a default value $\upsilon_2^{*} = 1$ for the missing estimate:
\begin{equation} \label{eq:upsilon_failure}
	\upsilon_2^{i^\prime} = \begin{cases}
	\ \upsilon_2^{*}&, \ for \ p = p_{fail} \\
	 \ \upsilon_2^i &, \ for \ p = 1 - p_{fail}
	\end{cases} 
\end{equation}
where $p_{fail}$ is the failure probability.

\subsection{Modelling noise}

The model adopted to simulate the effects of noise in the agents communication was the Additive White Gaussian Noise (AWGN). In real system simulations, noise is usually introduced in the communication channel modifying, in our case, the estimates of $\upsilon_2$ received from other agents. We model this process considering a noiseless communication channel that transmits an estimate already corrupted by noise when it enters it. Thus, let $\upsilon_2^{i^\prime}$ be the $\upsilon_2$ estimate made by agent $i$ in the presence of an independent and additive distributed noise $Z_i$. Thus:
\begin{equation} \label{eq:upsilon_noise}
	\upsilon_2^{i^\prime} = \upsilon_2^i + Z_i
\end{equation} 
where $Z_i \sim N(0,\eta)$ ($Z_i$ represents the \textit{gaussian noise}, that follows a normal distribution with $\mu = 0$ and $\sigma^2 = \eta$).

\subsection{Control Strategy in the Presence of Disturbances}
Let us consider a group of $N$ single integrator agents, whose dynamics is described by Eqs. \eqref{eq:control_law_decentralized}, \eqref{eq:u^c_final} and \eqref{eq:ud_formationcontrol}. The $\upsilon_2$ estimate calculated by agent $i$, in the presence of communication noise and failures can be expressed by:
\begin{equation} \label{eq:upsilon_perturbacoes}
	\upsilon_2^{i^\prime} = \begin{cases}
		\ \upsilon_2^{*} + Z_i &, \ for \ p = p_{fail} \\
		\ \upsilon_2^i + Z_i &, \ for \ p = 1 - p_{fail}
	\end{cases} 
\end{equation}

Consider $\lambda_2^{\prime}$ as the algebraic connectivity of a graph in the presence of disturbances and $\lambda_2^{i^{\prime}}$ as the estimate of $\lambda_2^{\prime}$ made by agent $i$. If the disturbance effect considered is assumed to be limited, (that is, there is no failure in \textit{any} communication between two nodes and noise is not \textit{infinite}), it can be concluded then that, similarly to what was seen in the absence of disturbances, $\lambda_2^i$ is a good estimate of both $\lambda_2^{\prime}$ and $\tilde{\lambda}_2^{\prime}$.

It can be noticed that the estimation error of $\lambda_2^{\prime}$ (the connectivity measure in the presence of failure and/or noise) is limited (that is, $\exists \Xi>0$ such that $\vert\lambda_2^{i^{\prime}}-\lambda_2\vert \leq \Xi, \forall i = 1, \ldots, N$). Likewise, it can be found that the difference between the connectivity estimates and the second smallest eigenvalue of $L$ is also limited (in other words, $\exists \Xi^{\prime} > 0$ such that $\vert \lambda_2^{i^{\prime}} - \tilde{\lambda}_2 \vert \leq \Xi^{\prime}, \forall i = 1, \ldots, N$). Hence, $\lambda_2^{\prime}$ can be used to locally estimate the real value of $\lambda_2$, inaccessible to the agents.

It is worth reminding that, besides not being possible to ensure that the estimation error of $\lambda_2^{\prime}$ tends to $0$, its value is larger as the effect of one or more types of disturbances in the communication betweeen agents gets larger.

\section{Matlab Simulations} \label{sec:matlab_simulations}
The results of the main simulations and experiments presented in this work are available \textit{online}\footnote{at \textit{goo.gl/HxIiYx}}.

A real-world robot system, both terrestrial (according to \cite{yan2006}) or aquatic (as in \cite{xiaobo2006}), is expected to have a control effort acting on a frequency approximately in the order of $10 \ rad/s$. Therefore, a \textit{first-order low pass filter} was added to the control signal with a transfer function given by Eq. \eqref{eq:tf_lp_filter}, attenuating the contribution of frequencies higher than $10 \ rad/s$ in $u^c$ in order to predict the performance of the proposed strategy in physical robots of the real world.

\begin{equation} \label{eq:tf_lp_filter}
	H(s) = \frac{10}{s + 10}
\end{equation}

A formation control problem with a varying number of agents ranging from $N = 3$ to $N = 10$ in an environment with $N_{obst} = 150 \ obstacles$ was considered for our experiments. Simulations have been carried out by considering the following parameter sets: $p_{fail} = \lbrace 0, 0.05, \ldots, 0.70 \rbrace$, $\eta = \lbrace 0, 0.1, 0.3, 0.5, 1.0, 5.0 \rbrace$.

For clarity reasons, the connectivity measure, its estimates and the control effort in the presence of disturbances (represented until now as $\lambda_2^\prime$, $\lambda_2^{i^\prime}$ and $u^{c^\prime}$) will only be referred to as $\lambda_2$, $\lambda_2^i$ and $u^c$. The estimate of a component of $\upsilon_2$ made by agent $i$ (denoted until now by $\tilde{\upsilon}_2^i$) will also be reported to simply as $\upsilon_2^i$ and for reference purposes, the connectivity measure in the absence of disturbances will be represented as $\bar{\lambda}_2$. 

The results presented consider a typical execution of the connectivity maintenance algorithm described in \cite{sabattini2012}. Typical runs of five agents performing formation control correspond to the robots starting at random initial positions and supposed to converge to a pentagonal configuration, while deviating from randomly placed point obstacles along the path.

\subsection{Failure}
\begin{figure}[!t]
\centering
\includegraphics[width=2.5in]{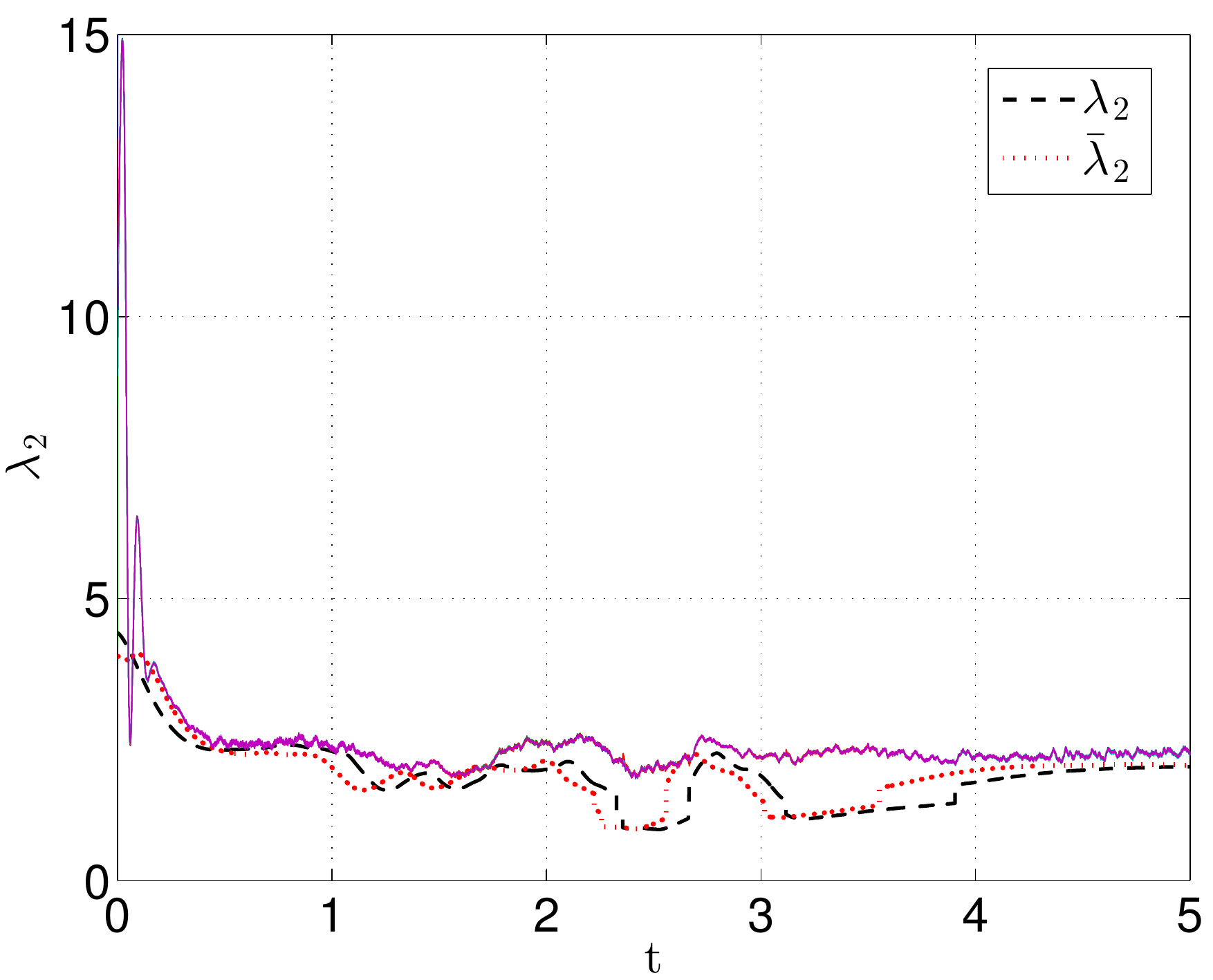}
\caption{$\lambda_2, \ \lambda_2^i \ and \ \bar{\lambda}_2\ for \ p_{fail} = 0.2 \ and \ \eta = 0 \ with \ N_{obst} = 150$}
\label{fig:lmbd_2-N5_f0p2n0_Nobst150_t5_1sim_lpFilt}
\end{figure}

In Figure \ref{fig:lmbd_2-N5_f0p2n0_Nobst150_t5_1sim_lpFilt}, the connectivity and its estimates evolution for $N = 5$ agents interacting with $p_{fail} = 0.20$ on an environment with $N_{obst} = 150$ obstacles is shown.

As can be seen, even in the presence of failures in the communication process, the system kept itself connected. The initial estimates ($\lambda_2^i \approx 10.17$ in $t = 0$) reach a maximum at the beginning ($\lambda_2^i \approx 14.9$ in $t \approx 0.02 \ s$) of the dynamics, while the connectivity decreases (from $\lambda_2 \approx 4.40$ in $t = 0$) and temporarily stabilizes at a specific level ($\lambda_2 \approx 2.32$ from $t_1 \approx 0.44 \ s $ till $t_2 \approx 1.07 \ s$), declining slowly as agents position themselves. The value of $\lambda_2$ changes because of the obstacle deviation process and varies from $\lambda_2 \approx 2.27$ in $t_{{obst}_0} \approx 1.07 \ s$, when agents start crossing the obstacles set, to $\lambda_2 \approx 1.68$ in $t_{{obst}_1} \approx 3.91 \ s$, when they leave the obstacle region, reaching a minimum of $\lambda_2^{min} \approx 0.91$ in $t_{min} \approx 2.53 \ s$. The simulation ends with $\lambda_2 \approx 2.01$ and $ \lambda_2^i \approx 2.26$ in $t = 5 \ s$.

As can be observed, adding failures to the communication process did not modify substantially the results observed in \cite{sabattini2012} regarding the connectivity dynamics, because both $\lambda_2$ and $\lambda_2^i$ kept themselves positive in the agent interaction. Compared to the results without any disturbances ($\bar{\lambda}_2$ in Fig. \ref{fig:lmbd_2-N5_f0p2n0_Nobst150_t5_1sim_lpFilt}), because of $p_{fail} > 0$, the values of $\lambda_2^i$ slightly deteriorate, but remain positive until the end of the simulation for $p_{fail} = 0.20$.

If the pseudorandom number generation (produced by the simulation environment) is approximated to a random one, then the failure probability is independent and equally distributed for \textit{all} agents. Consequently, the \textit{system failure} probability can be expressed by Eq. \eqref{eq:failprob_iteration}.

\begin{equation} \label{eq:failprob_iteration}
	p_{fail}^{sys} = N . p_{fail}
\end{equation}

In this case, $p_{fail}^{sys} = 5 \ . \ 0.2 = 1 \ fail / interaction$. That is, even with one robot failing at each interaction, on average the results observed in Fig. \ref{fig:lmbd_2-N5_f0p2n0_Nobst150_t5_1sim_lpFilt} show that the model responds well to this type of disturbance.

The evolution of $u^c$ regarding this dynamics can be seen in Figure \ref{fig:u^c-N5_f0p2n0_Nobst150_t5_1sim_lpFilt}. In Fig. \ref{fig:lmbd_2-N5_f0p2n0_Nobst150_t5_1sim_lpFilt}, it can be noticed that $\lambda_2^i$ begins by stipulating high values for the system connectivity, and then $u^c$ has low initial values. As the estimates diminish, $u^c$ rises and agents start crossing the obstacles region (which is done between $t_{{obst}_0} \approx 1.07 \ s$ and $t_{{obst}_1} \approx 3.90 \ s$), with a control effort variation between $u^{c^{min}} \approx 0.23$ in $t_1 \approx 1.91 \ s$ and $u^{c^{max}} \approx 0.90$ in $t_2 \approx 1.38 \ s$.

\begin{figure}[!t]
\centering
\includegraphics[width=2.5in]{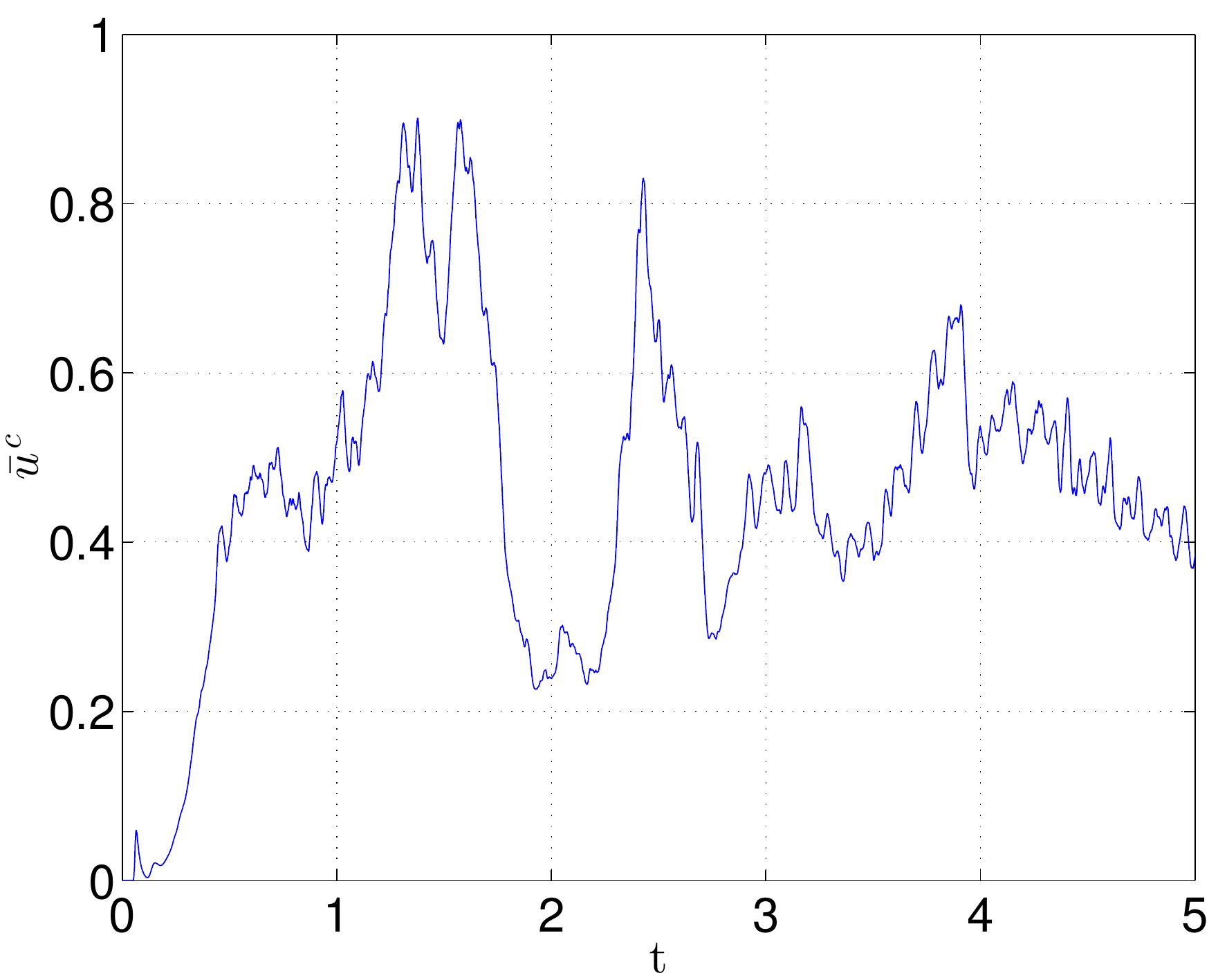}
\caption{$u^c \ for \ p_{fail} = 0.2 \ and \ \eta = 0 \ with \ N_{obst} = 150$}
\label{fig:u_c-N5_f0p2n0_Nobst150_t5_1sim_lpFilt}
\end{figure}

For $p_{fail} = 0.30$ (not shown here), results were qualitatively similar to the examined case, with connectivity maintenance for tipically-real controller agents. However, for $p_{fail} = 0.40$ connectivity was not maintained. In this situation, after reaching its maximum value, the estimates $\lambda_2^i$ fell and ended up becoming negative for some time, which made $\lambda_2$ (that was positive) decline and end up becoming zero in the sequence. Once the connectivity among the agents was lost, each one of them get incorrect values for the position of its peers outside its communication range. In this case, the estimates $\lambda_2^i$ lost their validity.

\subsection{Noise}

In Figure \ref{fig:lmbd_2-N5_f0d0n0,5_Nobst150_t5_1sim_Fs10k_lpFilt}, the values of the connectivity and its estimates are represented for agents communicating with $\eta = 0.5$ on an environment with $N_{obst} = 150 \ obstacles$.

\begin{figure}[!t]
\centering
\includegraphics[width=2.5in]{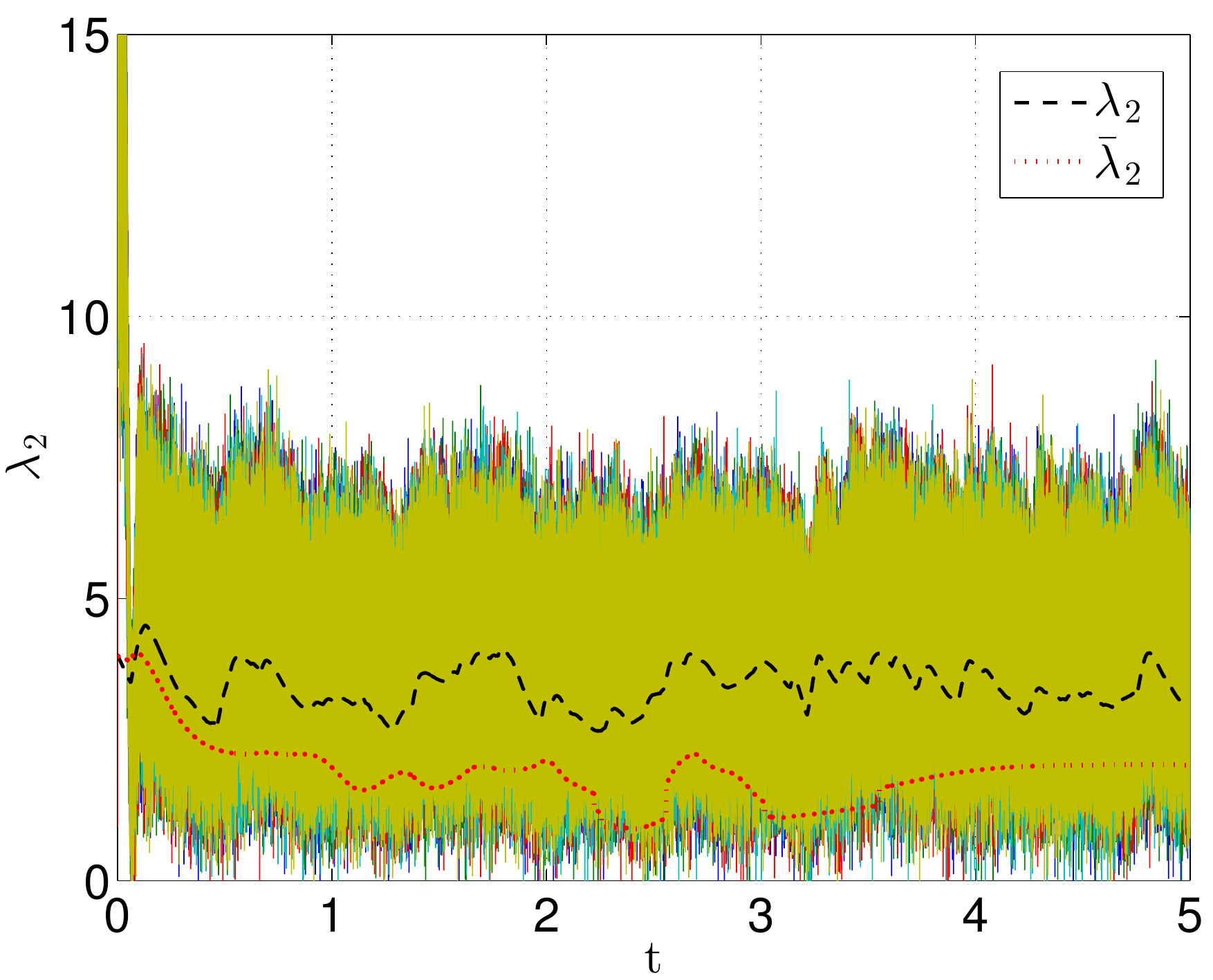}
\caption{$\lambda_2 \ and \ \lambda_2^i \ for \ \eta = 0.5 \ and \ p_{fail} = 0 \ with \ N_{obst} = 150$}
\label{fig:lmbd_2-N5_f0d0n0,5_Nobst150_t5_1sim_Fs10k_lpFilt}
\end{figure}

As can be seen, agents kept themselves connected even with noise in the communication process. The connectivity ($\lambda_2 \approx 4.00$ in $t = 0$) initially decreases and then grows until reaching its maximum ($\lambda_2 \approx 4.52$ in $t \approx 0.13 \ s$), while the estimates range between $\lambda_2^{i^{min}} \approx 0$ and $\lambda_2^{i^{max}} \approx 9.05$. The connectivity varies from $\lambda_2 \approx 3.32 $ in $t_{{obst}_0} \approx 1.04 \ s$ (when agents start going through the obstacles region) until $\lambda_2 \approx 3.88$ in $t_{{obst}_1} \approx 3.63 \ s$ (when they finish crossing it), with noticeable variations caused by the noisy connectivity estimates. Under the influence of noise, the connectivity maintenance control action is impaired and does not stabilizes at a specific value at the end of dynamics, as can be observed in the value of $\lambda_2$ in Figure \ref{fig:lmbd_2-N5_f0d0n0,5_Nobst150_t5_1sim_Fs10k_lpFilt}.

As noise is uniformly distributed, its influence on $\lambda_2^i$ ends up ``masking''  the agents temporary separation movement (and the consequent $\lambda_2$ descents) to the connectivity maintenance algorithm as the obstacles region is crossed. Therefore, $\lambda_2$ varies mostly as a consequence of noise throughout the interaction, and does not show any significant drop given by the obstacle deviation process (there is no valley in $\lambda_2$ between $t_{{obst}_0}$ and $t_{{obst}_1}$). Dynamic ends with $\lambda_2 \approx 3.12$ and $0.62 < \lambda_2^i <  6.89$ in $t = 5 \ s$.

Thus, communication noise did not affect substancially the connectivity dynamics previously obtained for the ideal-controller agents with no disturbances observed in \cite{sabattini2012}, despite harming the estimation procedure. Actually, agents just do not take exactly their own formation position during the dynamics, but oscilate around it, instead. 

\begin{figure}[!t]
\centering
\includegraphics[width=2.5in]{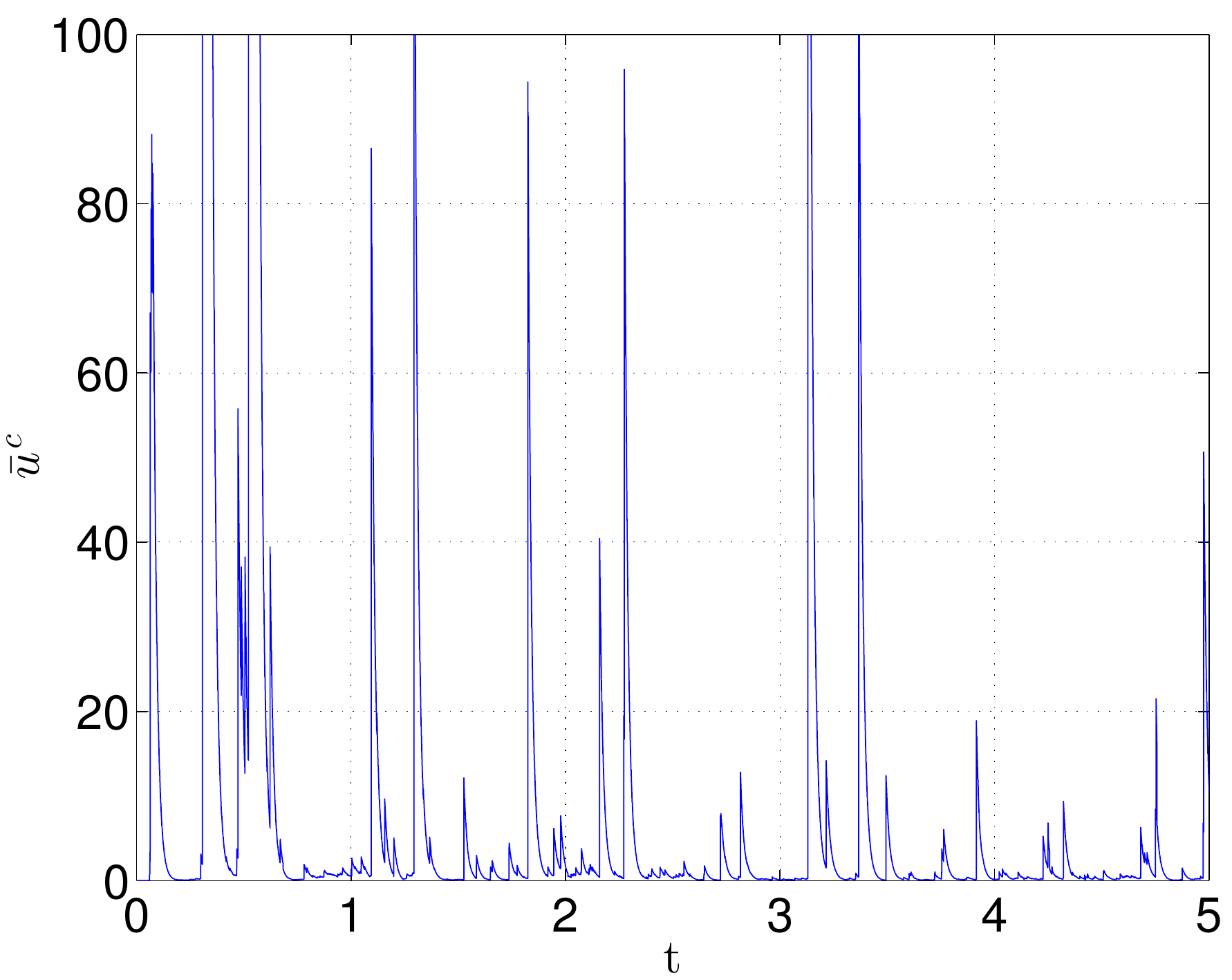}
\caption{$u^c \ for \ \eta = 0.5 \ and \ p_{fail} = 0 \ with \ N_{obst} = 150$}
\label{fig:u^c-N5_f0n0p5_Nobst150_t5_1sim_lpFilt}
\end{figure}

The evolution of $u^c$ regarding this dynamics can be found in Figure \ref{fig:u^c-N5_f0n0p5_Nobst150_t5_1sim_lpFilt}. 

In the beginning, a local maximum in the total control effort ($u^{c} \approx 88.2$ in $t \approx 0.07 \ s$) can be observed, followed by the global maximum ($u^c \approx 3.069$ in $t \approx 0.52 \ s$, in a not shown measure) as agents enter formation. Due to noise, agents poorly identify their own state and, as dynamic starts, they happen to reapproximate instead of taking their own formation position (as can be seen on the simulation videos referenced in the footnote) which explains the high value assumed by $u^c$ to put them in formation. Following that, the obstacles region start being traversed (between $t_{{obst}_0}$ and $t_{{obst}_1}$) and the value of $u^c$ decreases in intensity, because of the separation movement imposed by the obstacle avoidance action, but still demands local maxima in $u^c$ (as $u^c \approx 275$ in $t\approx 3.13 \ s$, for example) in order to keep the connectivity maintenance. Finally, after passing the obstacles, values assumed by $u^c$ decrease in intensity (with a local maximum of $u^c \approx 50.4$ in $t \approx 4.98 \ s$) but end dynamics taking local maxima to avoid disconnection in the presence of noise.

\begin{figure}[!t]
\centering
\includegraphics[width=2.5in]{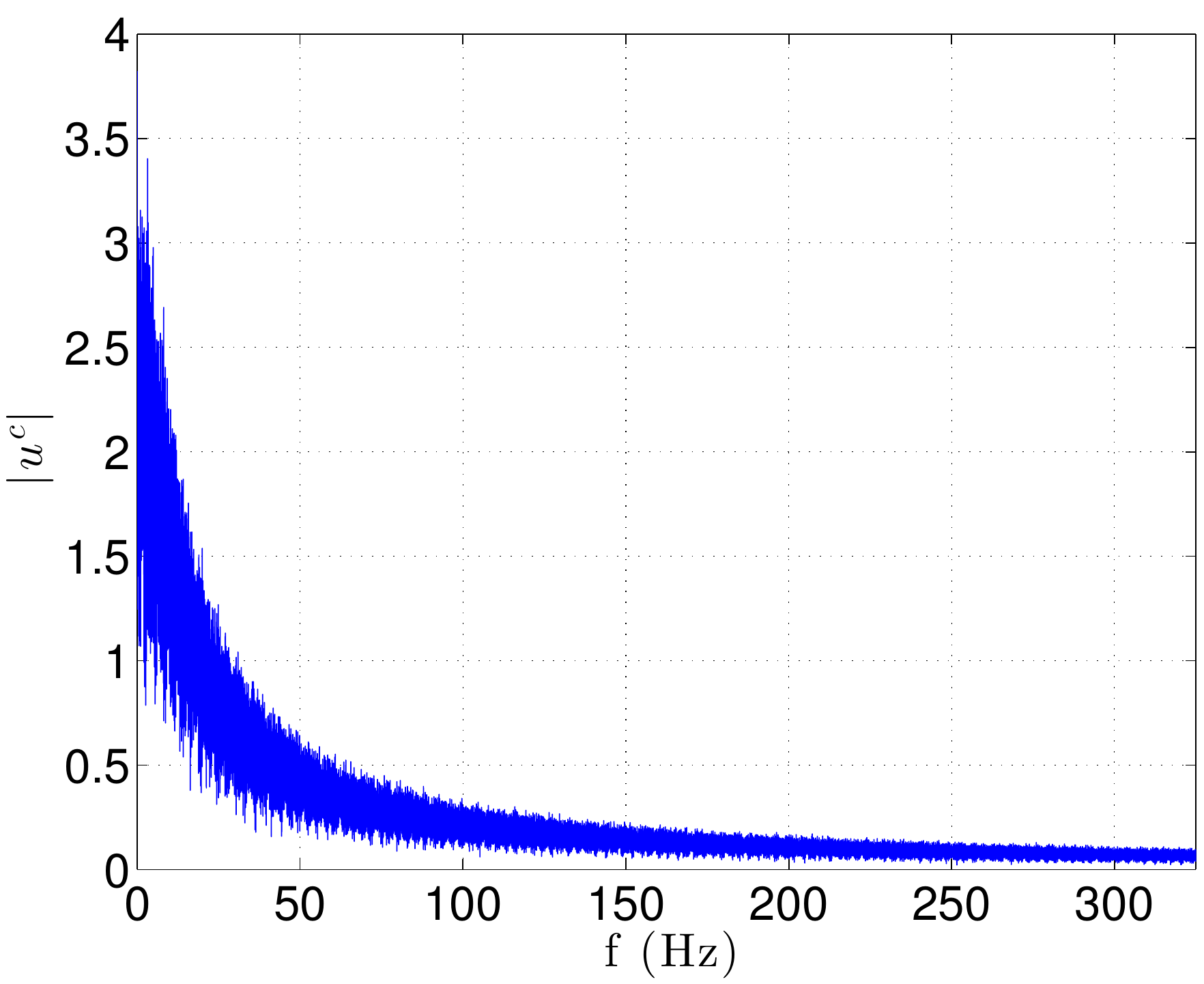}
\caption{$Frequency \ spectrum \ of \ u^c \ for \ \eta = 0.5 \ and \ p_{fail} = 0$}
\label{fig:u^c_x_f-N5_f0d0n0,5_Nobst150_t5_1sim_Fs10k_lpFilt}
\end{figure}

An analysis of the $u^c$ spectrum (see Fig. \ref{fig:u^c_x_f-N5_f0d0n0,5_Nobst150_t5_1sim_Fs10k_lpFilt}) shows that a system composed by tipically-real agents demands, for the connectivity maintenance in the presence of a noise rate $\eta > 0$, a control effort with an increased participation in high frequencies. In this case, these components are responsible for keeping the small amplitude variation around $\lambda_2$ during the obstables deviation in Fig. \ref{fig:lmbd_2-N5_f0d0n0,5_Nobst150_t5_1sim_Fs10k_lpFilt}. As $\eta = 0.5$, a noisy signal $Z_i \sim N(0,0.5)$ is added to the original information, what makes the values assumed by $\upsilon_2^{i^{\prime}}$ distribute themselves according to a normal distribution. As $u^c$ acts on a signal that changes more frequently (as consequence of noise), the presence of high frequency components in $u^c$ increases, as can be noted in Fig. \ref{fig:u^c_x_f-N5_f0d0n0,5_Nobst150_t5_1sim_Fs10k_lpFilt} for $f > 10 \ Hz$.

\section{Conclusion}
In this work, an analysis of the effect of different types of disturbances over a control algorithm was made, which through a decentralized estimation of the algebraic connectivity in a communication graph, guarantees the connectivity maintenance in a multi-robot system, given a random initial setup.

In the given analysis, the \textit{group connectivity was always kept for ideal agents} (in which there is an equitable response both for high and low requencies of the control effort, in results not shown). As can be seen in Section \ref{sec:matlab_simulations}, the high-frequencies performance can have in important role in guaranteeing the connectivity maintenance between tipically-real agents in the presence of high-levels of disturbance. In that case, analyses can be summarized as follows:

The model is failure-tolerant until a high specific $p_{fail}^{max}$ probability (where $p_{fail}^{max} < 40\%$). From this value, the $\upsilon_2$-estimates made by agents are not sufficient to ensure $\lambda_2 > 0$ while they move through a region with randomly-placed point obstacles. 

The model is noise-tolerant to every of the $\eta > 0$ cases analyzed, once this type of disturbance does not completely deteriorate the value of $\upsilon_2^{i^{\prime}}$ used in the estimation process of $\lambda_2^{i^\prime}$. Thereby, it is always possible for each agent, even though inaccurately, to compute $u^c$ in order to keep $\lambda_2 > 0$ if $\eta > 0$.

To put it briefly, as the connectivity maintenance is a necessary condition to the estimation procedure described in \cite{sabattini2012}, once it is lost, it is not possible to rely on the local estimates ($\lambda_2^i$) done by agents in the presence of high failure rates. However, in the real world, agents have inertial characteristics that prevent quick disconnections, and usually it is possible to recover from such situations by resetting the operation of the algorithm, for example.

Current work aims at proposing a strategy to smooth the disturbances impact in the global connectivity maintenance of a multi-robot system through the use of a Linear-Quadratic Estimator, filtering the communication noise in $\upsilon_2^{\prime}$ or decreasing the influence of communication failures in the agent interaction.




\section*{Acknowledgment}

The author would like to thank CAPES and Carlos H. C. Ribeiro thanks FAPESP (proc. nr. 2013/13447-3).




\begin{thebibliography}{1}


\bibitem{sabattini2011}
L.~Sabattini, N.~Chopra, C.~Secchi, \emph{On decentralized connectivity maintenance for mobile robotic systems.}, \hskip 1em plus
  0.5em minus 0.4em\relax CDC-ECC, Bologna, Italy, 2011

\bibitem{JiEgersted2007}
M.~Ji, and M.~Egerstedt,
\newblock \emph{Coordination Control of Multiagent Systems While
  Preserving Connectedness}.
\newblock IEEE Transactions on Robotics, 23(4), 693--703, 2007

\bibitem{hsieh2008}
M.~A.~Hsieh, A.~Cowley, R.~V.~Kumar, C.~J.~Taylor, \emph{Maintaining network connectivity and performance in robot teams}, \hskip 1em plus
  0.5em minus 0.4em\relax Journal of Field Robotics, Volume 25, Issue 1-2,  pages 111-131, January 2008
  
\bibitem{cao2010}
Y.~Cao, W.~Ren, \emph{Distributed coordinated tracking via a variable structure approach – part I: consensus tracking. part II: swarm tracking} \hskip 1em plus
  0.5em minus 0.4em\relax Proceedings of the American Control Conference, pp. 4744–4755, 2010
  
\bibitem{hollinger2010}
G.~Hollinger, S.~Singh, \emph{Multi-Robot Coordination with periodic connectivity}, \hskip 1em plus 0.5em minus 0.4em\relax IEEE International Conference on Robotics and Automation, May, 2010

\bibitem{yang10}
P.~Yang, R.~Freeman, G.~Gordon, K.~Lynch, S.~Srinivasa, R.~Sukthankar, \emph{Decentralized estimation and control of graph connectivity for mobile sensor networks.} \hskip 1em plus
  0.5em minus 0.4em\relax 2010

\bibitem{fiedler73}
M.~Fiedler, \emph{Algebraic connectivity of graphs}, \hskip 1em plus 0.5em minus 0.4em\relax Czechoslovak Mathematical Journal, 1973

\bibitem{degenaro06}
M.~C.~De~Genaro, A.~Jadbabaie, \emph{Decentralized Control of Connectivity for Multi-Agent Systems}, \hskip 1em plus 0.5em minus 0.4em\relax Proceedings of the IEEE International Conference on Decision and Control, page 3628, 2006

\bibitem{zavlanos09}
M.~M.~Zavlanos, H.~G.~Tanner, A.~Jadbabaie, G.~J.~Pappas, \emph{Hybrid control for connectivity preserving flocking}, \hskip 1em plus 0.5em minus 0.4em\relax IEEE Transactions on Automatic Control, 54:2869–2875, 2009. 

\bibitem{godsil2001}
C.~Godsil, G.~Royle, \emph{Algebraic Graph Theory} \hskip 1em plus
  0.5em minus 0.4em\relax Graduate Texts in Mathematics. Springer, 2001.
  
\bibitem{sabattini2012}
L.~Sabattini, C.~Secchi, N.~Chopra and A.~Gasparri, \emph{Distributed Global Connectivity Maintenance for Multi-Robot Systems}, \hskip 1em plus
  0.5em minus 0.4em\relax AUTOMATICA IT Congress, Benevento, Italy, 2012.   
  
\bibitem{Saber03consensusproblems}
R.~Saber and R.~Murray, \emph{Consensus Problems in Networks of Agents with Switching Topology and Time-Delays}, \hskip 1em plus
  0.5em minus 0.4em\relax 2003
  
\bibitem{fax04}
J.~Fax, R.~Murray, \emph{Information flow and cooperative control of vehicle formations.}  \hskip 1em plus
  0.5em minus 0.4em\relax IEEE Trans. Automat. Contr. 2004  
  

\bibitem{yan2006}
S.~Yan, F.~Zhang, Z.~Qin, S.~Wen, \emph{A 3-DOFs mobile robot driven by a piezoelectric actuator} \hskip 1em plus
  0.5em minus 0.4em\relax Smart Materials and Structures, 2006

\bibitem{xiaobo2006}
X.~Tan, D.~Kim, N.~Usher, D.~Laboy, J.~Jackson, A.~Kapetanovic, J.~Rapai, B.~Sabadus, Z.~Xin, \emph{An Autonomous Robotic Fish for Mobile Sensing}, \hskip 1em plus
  0.5em minus 0.4em\relax Intelligent Robots and Systems, IEEE/RSJ International Conference on, 2006

\bibitem{trefethen97}
L.~Trefethen, D.~Bau, \emph{Numerical Linear Algebra} \hskip 1em plus
  0.5em minus 0.4em\relax 1997

\bibitem{secchi2012}
C.~Secchi, L.~Sabattini, \emph{Decentralized global connectivity maintenance for interconnected lagrangian systems with communication delays} \hskip 1em plus
  0.5em minus 0.4em\relax Proceedings of the IFAC Workshop on Lagrangian and Hamiltonian Methods for Non Linear Control (LHMNLC), Bertinoro, Italy, 2012

\bibitem{secchi2013}
C.~Secchi, L.~Sabattini, C.~Fantuzzi, \emph{Decentralized global connectivity maintenance for interconnected Lagrangian systems in the presence of data corruption} \hskip 1em plus 0.5em minus 0.4em\relax European Journal of Control, Elsevier, 2013
  

\end{thebibliography}
%

\end{document}